\documentstyle[preprint,aps,floats,epsfig]{revtex}

\begin{document}
\tightenlines
\renewcommand{\thefootnote}{\fnsymbol{footnote}}
\preprint{\begin{tabular}{r}
KIAS-P00016 \\
YUMS 00-05 \\
hep-ph/0004155 \\
\end{tabular}}

\title{No CPT Violation from Tilted Brane 
       in  Neutral Meson$-$Antimeson Systems}

\author{G. Cveti\v c$^a$\footnote{cvetic@mail.apctp.org},~~ 
        S. ~K. Kang$^b$\footnote{skkang@kias.re.kr},~~ 
        C. ~S. Kim$^c$\footnote{
kim@kimcs.yonsei.ac.kr,  http://phya.yonsei.ac.kr/\~{}cskim/}~~
 and  ~~  K.~Lee$^b$\footnote{klee@kias.re.kr}}

\medskip

\address{$a:$ Asia Pacific Center for Theoretical Physics, 
Seoul 130-012, Korea\\
$b:$ Korea Institute for Advanced Study, Seoul 130-012, Korea\\
$c:$ Department of Physics and IPAP, Yonsei University, Seoul 120-749, Korea}

%

\maketitle

\renewcommand{\thefootnote}{\arabic{footnote}}

\begin{abstract}

Tilted brane in theories with large compact extra dimensions
leads to spontaneous symmetry breaking of the Lorentz and rotational
invariance in four dimensions, as shown by Dvali and Shifman. 
In this brief report, we point out that the mentioned Lorentz 
symmetry breaking, although leading to the CPT--violating
interaction terms, cannot lead to the CPT violation in the
experimentally interesting $K$--${\bar K}$ and analogous systems.\\
PACS number(s): 11.25.Mj, 11.30Er, 13.20Gd

\end{abstract}

\newpage

Recently, the authors of \cite{DS1} pointed out that
in theories with large compact extra dimensions
\cite{ADV} there exist solutions for the 3--brane
with a constant gradient energy -- tilted brane.
Further, they argued that such brane solutions,
via the change of the metric on the brane, 
lead in the four dimensions to interaction terms 
which violate the Lorentz and rotational invariance.

In this brief report, we show that such interaction
terms lead in principle to terms which violate the
CPT invariance. Subsequently, we argue that this
framework leads nonetheless to no CPT violation effects 
in the $K$--${\bar K}$ and analogous neutral meson--antimeson systems.

The tilted brane can be described by a
low energy effective theory. In this theory, 
the graviphoton ${\cal A}_{\mu}(x)$ obtains a mass
$m_{\cal A}\!\sim\!1 \ {\rm mm}^{-1}$ by eating up the 
corresponding Goldstone mode. After integrating
out the graviphoton, the effective theory for the other 
pseudo-Nambu-Goldstone boson(s) (pNGB) $\chi$ 
describing the dynamics of the brane
is represented by the four-dimensional Lagrangian density
\begin{equation}
{\cal L}_{\rm brane} = 
g^{\mu \nu} {\partial}_{\mu} {\chi}  {\partial}_{\nu} {\chi}  \ ,
\label{Lbrane}
\end{equation}
where $g_{\mu \nu}$ is the induced metric on the brane.
There are also fermionic terms, but they are not relevant
unless we include supersymmetry.
The tilted brane solution to the equation of motion
${\partial}^2 {\chi}\!=\!0$ is
\begin{equation}
\chi = \sqrt{T} {\alpha} x^j \ ,
\label{tbsol}
\end{equation}
where ${\hat x}^j$ is the direction along which the tilting occurs,
$T$ is the brane tension, i.e., the energy per
3--space unit volume, and $\alpha$ is the angle of the
tilting. The following relations hold: 
$T\!\sim\!M^4_{\rm P_f}$,
where $M_{\rm P_f}\!\stackrel{>}{\sim}\!1$ TeV 
($\sim$$10^{16}$ mm) is the
fundamental scale of gravity in the framework;
$\alpha\!\sim\!R H$, where $R$ is the typical size
of the extra dimension(s) ($R\stackrel{<}{\sim}1$ mm)
and $H^{-1}$ is the Hubble size ($\sim$$10^{28}$ mm).

The tiny tilting angle $\alpha$ causes the brane to be
``stretched'' by the factor of $1\!+\!{\alpha}^2/2$.
On the other hand,
${\alpha}\!=\!T^{-1/2} {\partial} {\chi}/{\partial} x^j$ 
by (\ref{tbsol}). This implies that the induced metric
$g_{\mu \nu}$ and the flat (untilted) metric
$g^{(0)}_{\mu \nu}$ are related
\begin{equation}
g_{\mu \nu} = g^{(0)}_{\mu \nu} + \frac{1}{2} T^{-1}
{\partial}_{\mu} {\chi} {\partial}_{\nu} {\chi} \ .
\label{indmetr}
\end{equation}
An analogous relation for the basis
4--vectors $e_a$ on the tilted brane
\begin{equation}
e_a^{\mu} = g^{(0) \mu}_a + \frac{1}{4} T^{-1}
{\partial}^{\mu} {\chi} {\partial}_{a} {\chi}
\label{eaind}
\end{equation}
follows from (\ref{indmetr}) due to $e_a \cdot e_b = g_{a b}$.
The kinetic term for the fermionic fields on the
tilted brane involves 
${\partial \llap /}_{\rm tilted} = e_a^{\mu} {\gamma}^a {\partial}_{\mu}$.
Therefore, when using (\ref{eaind}), the kinetic energy term in
the tilted brane background can be rewritten as
\begin{eqnarray}
{\cal L}_{\rm kin.} & = & 
({\overline \psi} {\partial \llap /} {\psi})_{\rm tilted}
= {\overline \psi} {\gamma}^{\mu} {\partial}_{\mu} {\psi}
+ \frac{1}{4} T^{-1}  \left( 
{\partial}^{\mu} {\chi} {\partial}_{\nu} {\chi} \right)
\left( {\overline \psi} {\gamma}^{\nu} {\partial}_{\mu} {\psi} \right) \ ,
\label{induced}
\end{eqnarray}
where all the derivatives are in the flat metric. 
If we now expand around the tilted brane solution (\ref{tbsol}), i.e., 
$\chi\!=\!\sqrt{T} {\alpha} x^j\!+\!\delta \chi$,
we obtain from the last term of (\ref{induced}) 
interaction terms which break the Lorentz and rotational
invariance
\begin{eqnarray}
\delta {\cal L} &=& \frac{1}{4} T^{-1/2} {\alpha}
{\partial}_{\nu} ( \delta \chi ) \left[ {\overline \psi}
\left( {\gamma}^{\nu} {\partial}^j\!+\!{\gamma}^j {\partial}^{\nu}
\right) \psi \right] + \frac{1}{4} {\alpha}^2 
{\overline \psi} {\gamma}^{\nu} {\partial}^{\mu} \psi 
{\big |}_{\nu=\mu=j} \ .
\label{liviol}
\end{eqnarray}
The first term ($\propto\!{\alpha}$), in addition, violates
CPT, because ${\partial}_{\nu} ( \delta \chi )$ is
odd and the term in $[ \ldots ]$ is even under CPT.
The violation of the Lorentz symmetry and of CPT
usually go hand in hand \cite{CK}. The last term in
(\ref{liviol}) is CPT--even.
 
The question immediately arising at this point is
whether the above CPT--violating interaction can lead
to CPT--violating effects in the $K$--${\bar K}$
or analogous neutral meson--antimeson systems. 
CPT violation phenomena
in the latter systems are in principle experimentally
detectable and were discussed in Ref.~\cite{KP}.
The current experimental bound on the CPT violation
in the $K$--${\bar K}$ system is already very stringent
\cite{bound}
\begin{equation}
\frac{ m_K - m_{\bar K} }{m_K}  < 2 \times 10^{-18} \ .
\label{dmK}
\end{equation}

As argued in Ref.~\cite{KP}, in string theory it is
possible to have a breaking of CPT that would lead
to nonzero energy shifts in the the quark inverse
propagators. If the 4--dimensional effective
theory arises from a string theory compactified at the
Planck scale $M_{\rm Pl}\!\sim\!10^{19}$ GeV,
this may lead, among other things, to 4--dimensional
effective interactions of the form \cite{KP}
\begin{equation}
{\cal L}_I \sim {\cal T} {\overline \psi} \Gamma \psi \ ,
\label{LI}
\end{equation}
where the field ${\cal T}$ is a CPT--odd 4--dimensional Lorentz
tensor, and $\Gamma$ denotes a gamma--matrix structure.
If tensor ${\cal T}$ acquires a nonzero vacuum expectation value
$\langle {\cal T} \rangle$, the interaction (\ref{LI}) 
generates a quadratic purely fermionic term and thus leads
to a tree level nonzero contribution $\triangle K$ to
the fermion inverse propagator.
An effective ``stringy'' interaction that would
lead to nonzero energy shifts in the quark inverse propagators is
\begin{equation}
\delta {\cal L} \propto \langle {\cal T}_{000} \rangle_0
{\overline \psi} {\gamma}^0 \psi \ ,
\label{stringy}
\end{equation}
where $\langle {\cal T} \rangle \propto 1/M_{\rm Pl}$.
The effective interaction (\ref{stringy})
is CPT--violating since ${\overline \psi} {\gamma}^0 \psi$
is CPT--odd. This term directly contributes to shifting
the energies of the quark inverse propagators --
for quark $q$ and antiquark ${\bar q}$ with opposite signs.
Therefore, such operator can contribute to nonzero
mass difference $m_K\!-\!m_{\bar K}$.

Now, let us check whether such a shift in the quark inverse
propagators can arise from the CPT--violating term in
(\ref{liviol}). That term still has one pNGB--field $\chi$ 
(of the tilted brane) present there beside the quadratic
quark field structure -- i.e., in contrast to (\ref{stringy})
it is not purely quadratic quark field term.
Therefore, it cannot contribute at the tree level to energy or
momentum shifts in the quark inverse propagators.
It can contribute to the quark inverse
propagators only through contraction of the $\chi$ fields,
which corresponds to the one--loop contribution mediated
by the $\chi$ field. However, this can only lead to
CPT--conserving phenomena because the CPT--odd term
of (\ref{liviol}) is then applied twice (in general: an even
number of times). Therefore,
the CPT--odd term (\ref{liviol}) in the tilted brane scenario
cannot lead to CPT--violating effects in the neutral
meson--antimeson systems, i.e., it cannot
induce nonzero mass difference
$m_K\!-\!m_{\bar K}$. The terms in (\ref{liviol}) may 
still lead to other CPT--violating phenomena in other
systems \cite{other}. 
However, the possible CPT violations in the $K$--${\bar K}$ 
system seem to be the most promising from the
experimental point of view -- because of the
very stringent experimental bound (\ref{dmK})
which will presumably be further refined.

In summary, we showed that the metric--induced terms
which violate Lorentz symmetry in the tilted brane
scenario by Dvali and Shifman, while being CPT--odd,
lead to no CPT--violating effects in the neutral
meson--antimeson systems.

\section*{Acknowledgement}
The work of G.C. was supported by the KOSEF Brainpool Program.
The work of C.S.K. was supported 
in part by  grant No. 1999-2-111-002-5 from the Interdisciplinary 
Research Program of the KOSEF,
in part by the BSRI Program, Ministry of Education, Project No.
99-015-DI0032, 
and in part by Sughak program of 
Korean Reasearch Foundation, Project No. 1997-011-D00015.
The work of K.L. was supported by the KOSEF 1998 Interdisciplinary
Research Program.

\end{document}